# Ransomware in Windows and Android Platforms

Abdulrahman Alzahrani, Ali Alshehri, Hani Alshahrani, Huirong Fu
Department of Computer Science and Engineering
Oakland University
Rochester, Michigan 48309
Email: faalzahrani, aaalshehri, hmalshahrani, fug@oakland.edu

*Abstract*—Malware proliferation and sophistication have drastically increased and evolved continuously. Recent indiscriminate ransomware victimizations have imposed critical needs of effective detection techniques to prevent damages. Therefore, ransomware has drawn attention among cyberspace researchers. This paper contributes a comprehensive overview of ransomware attacks and summarizes existing detection and prevention techniques in both Windows and Android platforms. Moreover, it highlights the strengths and shortcomings of those techniques and provides a comparison between them. Furthermore, it gives recommendations to users and system administrators.

## I. INTRODUCTION

Cyber-extorting malware is increasing and evolving continuously. Throughout its growth, it has posed major threats to modern technologies with daunting prevention tasks. For instance, computer systems' security critically depends on the ability of anti-malware products that must be abreast of new malware deployments. Authors of malware try to thwart anti-malware detection by implementing efforts that are more significant [1]. Moreover, some malware has a powerful instrument for illegal commercial activities such as ransomware.

Ransomware is defined as a form of malware that de-ceives users and infects their devices to limit or deny access. Hence, attackers demand payments from the victims for a promise to restore their data, and affected devices do not permit access until the ransom is paid [2]. Cyberattack writers have diversified their efforts to make money by using well-established methods. Ransomware has constituted as one of the most dangerous cyberattacks facing both organizations and individual users with global losses now likely running to billions of dollars. This type of attack is the latest trend that cybercriminals use for monetization by extorting their victims [3]. Moreover, ransomware's recent success has increased the appearance of new families in the last few years [4]. It spreads rapidly through websites, infected software and even email attachments.

As many categories of malware, ransomware uses some techniques to evade detection systems in order to trick the victims. It is able to encrypt files and establish secure con-nection with a Command and Control (C&C) server [5], [6]. Also, ransomware can exfiltrate users' information to a third party as well as perform multi-infection and process injection.

Ransomware exhibits behavioral differences if compared to traditional malware. For instance, most malware types aim to steal users' data, like banking credentials, without raising sus-picions. In contrast, ransomware operations behave differently since the attack notifies victims that their devices have been infected [7].

This paper focuses on Windows and Android, the plat-forms frequently targeted by ransomware attacks due to their popularity in both desktop and mobile market-share. The rest of this paper is organized as follows. Section II defines the concept of ransomware, and illustrates the functionality and how it works. Then, a brief overview of the history of ransomware cyberattacks is provided in Section III. Section IV lists the most affected industry sectors by malware attacks around the world. Next, Sections V and VI summarize the notable observed behaviors of ransomware in the Windows and Android platforms, respectively. Section VII provides a ransomware attack taxonomy, whereas Section VIII lists existing countermeasures and prevention techniques. Aiming to educate the users, Section IX provides a set of policies and recommendations. Finally, Section X concludes the paper.

## II. HOW RANSOMWARE WORKS

There are two types of ransomware: locker-ransomware and crypto-ransomware [8]. In the first type, blockers prevent victims from accessing their devices. Locker-ransomware can lock the whole screen of the device and demand payment. Typically, it masquerades as a notice from a local law en-forcement agency reporting an illegal action done by the user and indicating a spot-fine ransom. It does not encrypt any files or affect the data stored on the device. A notable example of locker-ransomware is WinLocker [9], [10].

By contrast, crypto-ransomware encrypts device services and files, or even the entire database that the system interacts with. Usually, once a file/folder is encrypted, it will be deleted from the device. Then a threatening message is displayed with a link to the acceptable payment methods and instructions [11]. Encryption keys can be generated locally by the ransomware on infected machines, or remotely on a C&C server as il-lustrated in subsection V-C. Furthermore, accessing encrypted files is restricted by withholding the decryption key [5]. A notable example of this type is the CryptoLocker family [12].

After performing a successful infection, the malicious pro-gram notifies the user by displaying a ransom message, which relies on Bitcoins currency for more confidentiality and anonymity. This message contains instructions on how to pay the ransom in order to restore access to the encrypted data. Those persistent messages can be generated by calling some dedicated API functions such as CreateDesktop(), which creates a new desktop and makes it a default for infected machines, locking out the users [13].

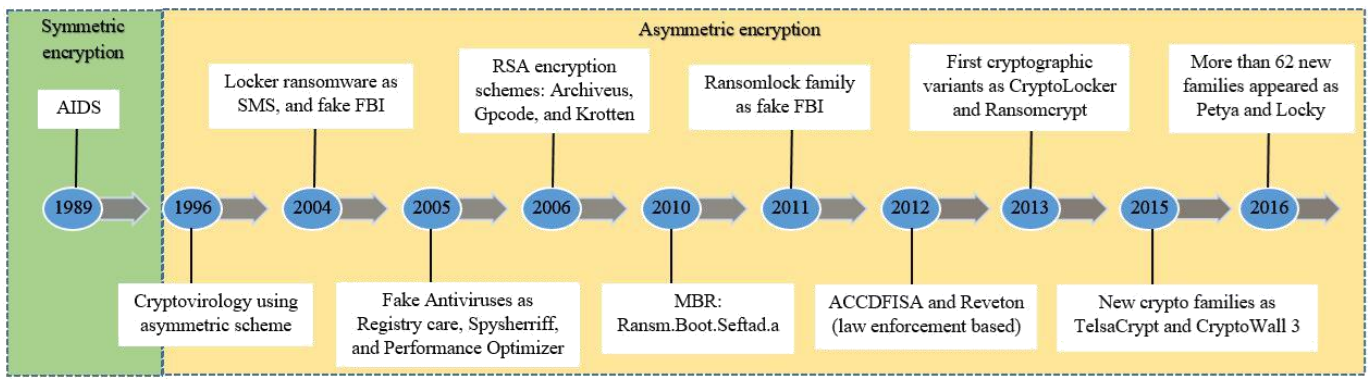

Figure 1: Ransomware timeline over the history

Additionally, ransomware cybercriminals do not deter large due to their extortion campaigns. Hence, analysts discriminate ransomware from wiping attacks. Ransomware attackers are mostly on the benefit of an unlikely trusted relationship with infected users. Attackers rely on the tenet that they will abide by an implicit agreement with their victims to restore access to their files once they make the payment. However, "skiddie" cybercriminals trick their victims into paying the ransom and never return the victims' files [14].

### III. HISTORY OF RANSOMWARE

Crypto-based attacks are relatively old. The first known form was written by Joseph Popp and released in 1989 named AIDS, short for Aids Info Desk. It is also known as PC Cyborg, which was the first crypto ransomware at-tack seen in December 1989 [15]. This version replaced the AUUTOEXEC.BAT file, and counted how many times the machine was booted. Once it reached a certain time, it hid directories and encrypted their names. Then it displayed a message claiming that a license of some software that the user used is expired, and asked the user to pay some amount in order to obtain a repair tool. This attack used a combination of a symmetric cryptography and an initialization vector to encrypt files in a victim's machine. Hence, the key could be extracted from its code [11].

Since then, AIDS has been present until now, but has significant changes [16]. Adam L.Young and Moti Yung proposed the concept of using asymmetric key in 1996 [17]. At that time, cryptographic libraries were restricted by governments' legislation. They showed that the AIDS was ineffective by using one key. Therefore, they introduced public key cryptography for such attacks. In this attack release, the user's files were encrypted using the ransomware author's public key. In order to decrypt those files, victim needs to get a decryption key, which could be obtained after they pay their ransoms.

Locker ransomware appeared as SMS, the master boot record (MBR), and fake FBI ransomware. The first locker ransomware came into existence in 2004 as Antivirus software. In 2005, a series of fake Antivirus ransomware variants appeared such as Spysherriff, Registry care, and Performance Optimizer [18]. The last two variants offered paid solutions for problems, which did not actually exist in the victim's machine. Furthermore, they were deployed over the Internet until 2008.

Cybercriminals have implied more effective and sophisticated Trojans using asymmetric RSA encryption schemes. A pair of keys are generated in asymmetric cryptosystem known as public key and private key. The public key is embedded in the payload of the ransomware to encrypt data on the victim's machine. On the other hand, the private key is kept secret and only known by the payload's writer. Thus, encrypted data can be recovered at the writer-side, and malware analysts cannot determine the private key from monitoring the operations of ransomware as used with symmetric schemes. Notable examples of this type are Gpcode, MayArchive, TROJ.RANSOM.A, Archiveus, and Krotten [19].

New families appeared and started spreading in 2006 such as Cryzip and Archiveus [20]. These variants sniff out for specific type of files to make them inaccessible. For instance, Cryzip encrypts particular extensions, then moves those encrypted files into a zipped folder. Whereas Archiveus moves files into a password protected folder.

In 2010, MBR ransomware variants made their first appearance under name Ransom.Boot.Seftad.a. Another type called bootlock.B came into existence in 2011. It replaces the original MBR with its code and locks the user's machine then displays a ransom message at booting time. This type never uses encryption. Moreover, the year of 2011 was the first year of Fake FBI ransomware to come into existence by the appearance of Ransomlock family [21], [22]. Additionally, in 2012 other families of the Fake FBI ransomware started spreading such as ACCDFISA and Reveton. The ransom payments of these families were displayed in an official format as a local law enforcement agency.

In 2013, new families appeared including Virlock and Kovter with the continuation of new variants of Reveton and Ransomlock families. A huge comeback was made by crypto ransomware with Cryptolocker, Cryptolocker 2, Dirty Decrypt, Crilock, and Ransomcrypt [13]. Whereas in 2015, new classes of Cryptolocker, TelsaCrypt, Cryptoblocker, CryptoTorLocker, CryptoFortress, Ransomcrypt, Ransomweb, Vaultcrypt, Troldesh, Pclock, Cryptowall 3 and Cryptowall 4 were introduced [8]. Most of recent crypto variants grow with sophisticated and diverse encryption techniques. They are written in scripting languages, becoming more targeted and exploiting new infection paths. For instance, Cryptowall 3, which has been released into the wild by a Russian cybercrime gang, uses Tor anonymity network for C&C communication [5], [23]. Ransomware-as-a-service came to existence, which makes ransomware attacks available to everyone receiving a commission on every successful campaign [20].

Table I: Top five industry sectors targeted by ransomware

| Sector Name | Percentage of attacks |
|---|---|
| Education | 23% |
| IT Telecoms | 22% |
| Financial Services | 21% |
| Government Sectors | 18% |
| Healthcare | 16% |

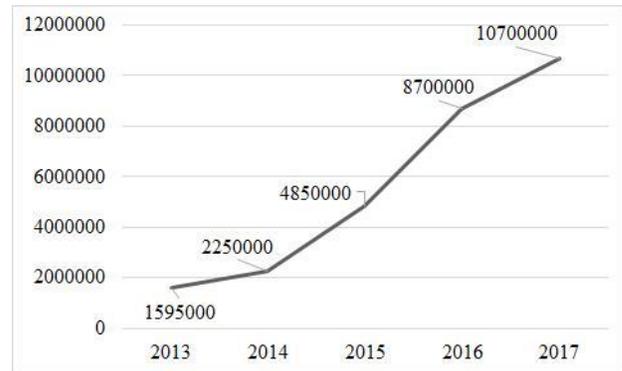

Figure 2: The recent growth of ransomware attacks

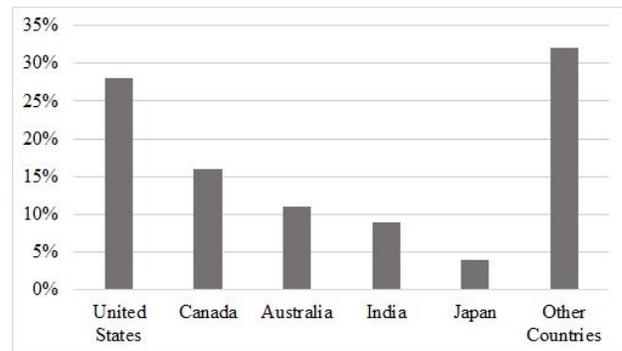

Figure 3: Top five countries hit by ransomware

According to Kaspersky report for the year of 2016 [14], there were 62 new ransomware families appeared. It can be declared the year of ransomware. Ransom32, Locky, PH-PRamson and HydraCrypto are notable examples of those variants. However, starting from the year of 2016, security organizations have begun a union to fight back under name No More Ransom. This collaboration has resulted in a number of free decryption tools that have helped thousands of users to recover their data. In addition, the biggest surprise of this year was the shutdown of TeslaCrypt by the malware actors themselves. Figure 1 summarizes the ransomware history timeline.

IV. RANSOMWARE VS. BUSINESS

Many ransomware attacks are indiscriminate and the infection is similar for businesses and individual consumers. However, a number of cybercriminal groups have begun implementing specific ransomware attacks to target businesses and infect multiple computers on a single network. Further-more, major ransomware robbers are capable of pushing their malicious activities to millions of computers around the globe. The perfection of this business model has created a competitive mentality among attackers. As a result, the infection numbers are trending upwards, and victims find their valuable data locked with strong and unbreakable encryption [24]. Further-more, according to the annual report by Kaspersky, by the end of 2016, there was one attack every 40 seconds targeting business organizations. In addition, one in five small and medium-sized business (SMBs) have paid their ransoms, but never received their data back [4].

Another report was made by Symantec in 2016 [25], an incremental number of gangs have focused on attacking large organizations. Those attacks have used techniques that are commonly seen in cyberespionage campaigns and involved a high level of technical expertise to break into targeted networks. Once it traverses the network successfully, it can cause massive operational disruptions or even stops the entire business, leading into serious damages to revenues and reputa-tion. Table I shows the most common industry sectors that are hit by ransomware. Additionally, the curve in Figure 2 shows the tremendous spike of this type of malware in the last five years until September 2017. Whereas, Figure 3 shows the top five countries that were hit by ransomware attacks as of the end of 2016 [25].

V. RANSOMWARE ANALYSIS ON WINDOWS PLATFORM

Sectors' data can be infected through compromised software, malicious email attachments, and drive-by download exploit kits and advertisements, which is installed without user knowledge when browsing suspicious websites [26], [27]. The following list contains the notable observed behaviors of ransomware in the Windows platform:

A. File System Activities

A large number of malware sample executions lead to file system changes. During the execution of ransomware payload, new files are created and existing files are modified or even deleted. Files such as .txt,.log,and.tmp are usually created and modified constantly [28]. For instance, CryptoWall variants modify PIPE\lsarpc,MousePointManager as well as an .exe file inside the temp folder that belongs to the administrator account. Also, they modify system.pif in the Start Menu in order to restart some particular software even if the machine is rebooted [16]. Additionally, some Cryptolocker variants create some executable files and mod-ify folders under C:\DocumentsandSettings as well as TemporaryInternetFiles in order to change the browser's homepage and display the ransom message [29]. Moreover, Ransomware variants can delete all volume shadow copies, back-up files, and restore points by using the vssadmin tool [30].

### B. Registry Activities

Registry is a hierarchical database that stores low-level settings and operations for the operating system. It can be used by the kernel, services, device drivers, the Security Accounts Manager (SAM), and user interfaces (UIs). Registry contains settings, options, and other configuration information for software and hardware components that have been installed on the OS. Some malware (e.g. ransomware variants) creates a registry key once it is installed to take control over the system. For instance, Windows\CurrentVersion\Policies\ System key is used to prevent users from invoking the task manager [28], [1].

Furthermore, most types of ransomware families manipulate or delete many existing registry keys, subkeys, values, and value data in order to persist and function properly. Notable examples of those registry keys can be HKEY_LOCAL_MACHINE\SOFTWARE\Microsoft\ Windows\CurrentVersion\Run,HKEY_LOCAL_ MACHINE\SOFTWARE\Microsoft\WindowsNT\ CurrentVersion\WinLogon, and HKEY_LOCAL_ MACHINE\System\CurrentControlSet\Control\ Nls\ComputerName\ActiveComputerName. However, some others read registry values, such as Microsoft Strong Cryptographic Provider (MS STRONG PROV) – HKLM\Software\Microsoft\Cryptography\ Defaults\Providerv1.0, which is used as the default RSA Full (PROV RSA FULL) cryptographic service provider (CSP) [31].

### C. Network Activities

Moving from symmetric encryption to asymmetric encryption in cyber-extoring threats has enhanced more communications through the network. In a ransomware scenario, when a machine is infected, it communicates with the C&C server to obtain a public key. This can be done through multiple proxy servers that are typically hacked, as shown in Figure 4. Most of these communications are made through ports 80 and 443 (TCP connections), and port 53 (UDP connections) [16]. For instance, CryptoWall variants use HTTP protocol (POST messages) to contact a C&C server. Instead of using IP addresses directly, they use domain names [5]. In addition, some ransomware variants have the ability to crawl the entire network to encrypt all existing files and lock all attached computers [9].

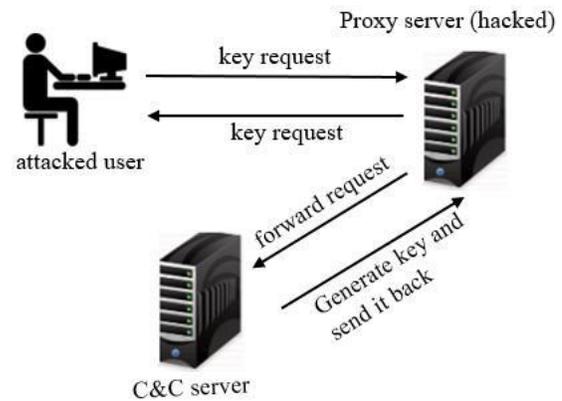

Figure 4: Asymmetric encryption scheme

### D. Communication Devices

Variants of cyberattacks use some devices' controls during their executions on the endpoint machines. Cryptlocker, Reve-ton, and CrypCTB ransomware families use many devices to communicate, such as:

\Device\Afd\Endpoint, which is a symbolic link referred to the device that transfers packets to the local network and Internet.

\Device\Tcp, for TCP connections.

\Device\KsecDD provides the kernel security device driver.

MountPointManager driver is responsible for main- taining persistent drive letters and names for volumes.

The rest of the ransomware variants use \Device\ KsecDD device to communicate with the victim's machine [32], [33].

### E. Encryption Mechanism

Generally, a cryptosystem and its suites are defensive, and provide privacy and security [17]. It is a boon to the secu-rity society. However, it makes a ladder for extortion-based cryptovirus attacks. Some techniques of cryptosystems make use of standard Windows functions to perform encryption such as CryptoWall and CryptoLocker families, by simply calling CryptEncrypt [13].

Modern ransomware mixes techniques from well-established benign cryptography suites called hybrid cryptosystem. In this technique, the ransomware generates a random symmetric key (commonly referred to as a session key) for each targeted object (message, file, folder, etc.), and encrypts it by using its key. Subsequently, the hybrid cryptosystem encrypts the symmetric key with an asymmetric encryption using a public key generated on the attacker's command and control infrastructure and embedded to the payload infecting victims' machines [13]. Thus, the asymmetric cryptographic operations are only required to encrypt and decrypt the small symmetric key regardless of the encrypted content's size. As a result, the strength of ransomware racket is equal to the security of a hybrid cryptosystem [34].

Table II: Common C&C commands

| Command | Usage |
|---|---|
| COMMON_HELLO | Checks if encryption key matches received key |
| COMMAND_SECRET | Changes communication encryption key |
| COMMAND_BOT_ID | Sets new BOT_ID malware |
| COMMAND_ENCRYPT | Encrypts files on external storage |
| COMMAND_DECRYPT | Decrypts files on external storage |
| COMMAND_PASSWORD | Sets new password used to encrypt/decrypt files |
| COMMAND_JID_CONFIG | Modifies the XMPP accounts parameter BUILTIN_JID |
| COMMAND_SERVER_MESSAGES | Instructs malware to change all COMMANDS keys to new ones |
| COMMAND_VOUCHER_MESSAGE | Sets a VOUCHER ERROR MESSAGE |
| COMMAND_CALL | Calls given number |
| COMMAND_SMS | Sends SMS to given number |

### F. Locking Mechanism

Types of ransomware in this mechanism lock out machine resources as well as input devices connected to the infected machine. In some cases, they use JavaScript codes to change the settings of browsers such as Google Chrome, or to flash a full screen image to block all other windows. Then it creates a fake message claiming a prohibited action has been done by the user, and includes a ransom note. This approach keeps the system and its files untouched, which makes it possible to remove such a malware by restoring the system to its original state. Moreover, it allows a limited functionality access such as enabling the mouse and numeric keyboard keys on the victim's machine to only interact with the ransomware, and denies any other access [35].

## VI. RANSOMWARE ANALYSIS ON ANDROID PLATFORM

In most cases, reverse engineering technique is used to analyze this class of Android malware. Mainly, it focuses on the Manifestfile.xml and the course code files of the app. The following subsections describe the most malicious payload activities that are observed by existing analysis tools.

### A. Privilege Escalation

Sensitive operating system services require special privilege and access control to execute their tasks in secure manners. Services that are running on computers and connected to the Internet present a target for adversaries to compromise them [36]. As a result, it can lead to unauthorized access to some sensitive resources. Configuration oversight and design flaw can lead to a privilege escalation. For example, programming errors in privileged services may allow adversaries to compromise the system in the form of unauthorized acquisition of privileges [37]. Android malware authors pack and obfuscate their payload to bypass restrictions imposed by sandboxes. In addition, if attacks obtain root privileges, they can break down the whole security system and pose serious threats, which makes removing malicious applications from the device difficult [38].

### B. Remote Control

As any other regular piece of software that requires secure communication, earlier malware packages communicate with a website through HTTPS to get encryption keys. Typically, when an application makes a secure HTTP request to communicate with a suspicious target, it is a clear hint of malicious activities [39]. However, the new variants communicate via the Extensible Massaging and Presence Protocol (XMPP). Such a protocol facilitates communications with the C&C server, which looks like normal instant message (IM) communica-tions. Messages can be encrypted by using the Transport Layer Security (TLS). Thus, it is more difficult to detect ransomware using anti-malware software. Table II summarizes the common commands used in C&C server communications.

### C. Sensitive Data Collection

Android operating system APIs provide installed apps with large amounts of user's information. This information can be locations, contacts, IMEI, call logs, profile, browser history and bookmarks, phone state, SMS, installed apps, etc. Mal-ware uses collected information for different purposes without the user's awareness [40]. Additionally, ransomware variants check the running tasks on the device in order to evade and bypass detection systems [41]. With some special permissions and components, ransomware apps can manipulate and kill any other running processes that are not the malware itself [42].

### D. Encryption and Locking Mechanisms

Historically, the potential use of cryptographic schemes for offensive purposes was documented for decades, as Section II indicated. Recent ransomware attacks try to obtain administrator privileges in order to perform their activities such as setting a new PIN for the screen and locking the entire device [16]. Another locking technique is to superimpose a full screen alert message so that the user can only interact with the ransomware dialog. Furthermore, trapping key-pressure events is another common locking way used by some variants to deny switching away from the lock screen [43].

Table III: Permissions requested by ransomware

| Permission | Behavior |
|---|---|
| READ_PHONE_STATE | Permits to access phone state |
| INTERNET | Permits apps to connect to the Internet |
| READ_HISTORY_BOOKMARKS | Permits apps to read browser history and bookmarks |
| BIND_DEVICE_ADMIN | Permits interaction between device administration receiver and the system |
| ACCESS_NETWORK_STATE | Permits apps to access the network information |
| ACCESS_WIFI_STATE | Permits apps to access Wi-Fi networks information |
| WRITE_EXTERNAL_STORAGE | Permits apps to write to external storage |
| READ_EXTERNAL_STORAGE | Permits apps to read from external storage |
| WRITE_SETTINGS | Permits apps to read or write system setting |
| SYSTEM_ALERT_WINDOW | Permits apps to alert system |
| RECEIVE_BOOT_COMPLETED | Permits apps to receive the ACTION_BOOT_COMPLETED that is broadcasted after the system finishes booting |
| ACCESS_COARSE_LOCATION | Permits apps to access approximate location |
| ACCESS_FINE_LOCATION | Permits an app to access precise location |
| WAKE_LOCK | Used to prevent the device from going to sleep |
| KILL_BACKGROUND_PROCESSES | Used to kill any other running processes that are not the malware itself or the phone setting application |
| INSTALL_SHORTCUT | Permits an app to install a shortcut in Launcher |
| GET_TASKS | Permits an app to get information about the currently or recently running tasks |
| GET_ACCOUNTS | Permits access to the list of accounts in Accounts Service |
| READ_CONTACTS | Allows apps to read contacts information |
| READ_CALL_LOG | Allows apps to read call log |
| DISABLE_KEYGUARD | Used to disable keyguard when it is not secure |
| CALL_PHONE | Permits an app to initiate a phone call without going through the Dialer UI for the user to confirm the call |
| READ_SMS | Allows apps to read SMS |
| RECEIVE_SMS | Allows apps to receive SMS |
| SEND_SMS | Allows apps to send SMS |
| CAMERA | Permits apps to access the camera |

Ransomware variants imply various extortion techniques in order to encrypt the data and lock the device. Usually, it searches for particular files to encrypt them. For example, Crypto-ransom like Pletor and Simplocker use AES encryption scheme to encrypt data that is presented in SD card. Further-more, some variants only encrypt the device and leave the data untouched. Whereas some others encrypt the data and leave the device open. Thus, even after removing the malware, the victim has no choice than paying the ransom in order to restore access [43], [44].

E. Permissions Used

An app installation process demands some permissions to be granted in order to function properly [45]. Users can see permission requirements prior and after the installation. However, ordinary users may unintentionally download some apps without paying attention to their permissions. Such a security weakness allows malware to pretend as normal apps [46]. Table III shows common permissions that a ransomware variant may request. Note that some permissions can be demanded by benign apps as well [47].

VII. RANSOMWARE TAXONOMY

So far, this article has done a systematic review of the terms related to ransomware attacks and summarized behavioral descriptions of the topmost families based on the number of infected users. It categorized those families into two main categories based on the variant actions and behaviors. These categories are cryptographic and locking ransomware as fol-lows:

A. Cryptographic Ransomware

Ransomware scenarios have been used for mass extortion. However, a pronounced trend in recent years has been the shift towards cryptographic ransomware. The proportion of new crypto-ransomware variants is growing every year. The growth of this type can be explained by the fact that it is the most effective form of ransomware [25]. A few years ago, the market was dominated with misleading apps. Many of those applications were designed to pose as an anti-virus software [48]. They inform users that there is something wrong with their machines, which is a result of the malware infection, and then they request some amount of money in order to fix the problem.

Traditionally, successful ransomware attacks perform one or more of the following activities:

Indiscriminate encryption: this crypto-type aggressively encrypts and deletes the user's private files. Cybercriminals can overwrite files with the encrypted versions, or delete the original files by using Windows API functions or Windows Secure Deletion API to perform secure deletion. A notable example of this type is TorrentLocker, which encrypts the first two megabytes of all existing files on the system [49], [7].

Selective encryption: this form of cryptovirus attacks, encrypts, and deletes user's private files based on specific attributes such as size and extensions. It performs a selective encryption in order to avoid detection systems [16], [7]. A notable example of this type is any variant of the Cryptowall family [50].

The following list summarizes topmost cryptographic ran-somware families. It aims to define each type and how it works:

1) Petya: Petya appeared in March 2016 as a partial disk cryptosystem. Instead of file encryption, this variant encrypts the master file table. Therefore, files on disk will be prevented from being located. Once the ransom is made, the encryption key will unlock the master file table and reboot record, and then the malware boot loader will be removed. Petya variant is especially troublesome because unlike other variants, it does not require any Internet connection in order to generate the encryption key [23], [51].

A new variant has appeared in June 2017 called NotPetya. Petya was described as a criminal enterprise for making money. However, this latest form is not designed to make money. The main goal behind NotPetya is to spread fast and cause damage with a plausibly deniable cover of ransomware. It takes out businesses from shipping ports and supermarkets to ad agencies and law firms. Furthermore, once this well-oiled destructive program infects a corporate network, it worms its way from one computer to another and harms the file systems of the infected machines. After that, it demands about $300 in Bitcoin to unscramble the hostage data. In addition, NotPetya contains certain mechanisms to collect this money from victims and exchange decryption keys [52]. The following is a summary of the NotPetya outbreak:

> It uses other tools to spread through a network and infects machines such as a tweaked build of open-source Minikatz, which is used to extract network administrator credentials out of the machine's running memory. Collected information is used to communicate with other machines by using PsExec and Windows Management Instrumentation Command-line (WMIC) to infect them and execute commands. The PsExec is a light-weight telnet-replacement that allows cybercriminals to execute processes on other systems. It can also identify known hosts by using the Dynamic Host Configuration Protocol (DHCP) service.
> A stolen and leaked version of the NSA EternalBlue SMB exploit as well as the agency EternalRomance SMB exploit are used to inject malicious codes into other systems.

As of March 2017, Microsoft has patched these cyber-weapons attack vulnerabilities. However, the credential theft becomes more successful at places that are on top of their Win-dows updates. As a simple solution against these new attacks, researchers explained that before overwriting the computer's Master Boot Record, the ransomware checks for the perfc file in the C:\Windows folder. If it does not exist, then the ransomware will encrypt the computer. However, if that file is present, then the ransomware will stop. Hence, in order to halt the encryption in its tracks, users need to create a perfc file in the C:\Windows folder and make it read only [51], [52].

2) Simplocker: Simplocker a mobile Trojan form discovered in June 2014 and considered as the first type of encrypt-ransomware to attack Android devices. At that time, Simplocker (or Simplelocker) was heavily inspired by desktop crypto-ransomware. However, the Android's security model was able to curtail its scam. The reason behind that was related to the security restriction, which prevents apps from accessing files and data that belong to other apps unless permissions are granted to do so. Android.Simplocker uses 256-bit AES key (symmetric encryption) to encrypt files on the user's device. The key is included in the application code, which means that it does not need to communicate with a C&C server to complete its encryption process.

However, malware writers can send commands to this type through SMS messages such as encrypt/decrypt a user's file. Additionally, in previous versions of Android operation systems, files such as images and media were stored on unprotected external SD memory cards. Hence, they could be accessed by other applications. Malware applications, like Simplocker, could access those files stored in the memory card and encrypt them [18].

3) GPcode: GPcode a Trojan malware that encrypts files with certain extensions (such as .html, .rar, .txt, .doc, .jpg) on the infected machine or drives and asks users to contact its author to buy a decryption solution and retrieve access. GPcode family (GPC) was reported as the first wave of modern ransomware, which started in May 2005 with Trojan.Gpcoder as one of the crypto ransomware threats [18]. Initially, variants of Trojan.Gpcoder used custom-encryption techniques that were weak and easy to beak. Despite those initial failures, the authors of cyberattacks continued to enhance better tech-niques and create newer versions of GPcode threats such as backdoors. Furthermore, a new version was discovered in November 2010 that used a stronger encryption technique (RSA-1024 and AES-256) and overwrites the encrypted files to make file recovery nearly impossible [53].

In addition, once a variant is installed on the machine, it usually creates two registry keys. The first registry key is used to ensure that it is running on every system boot. On the other hand, the second registry key monitors the Trojan's progress in the infected machine and count the number of files that have been analyzed by the malicious code. Recent variants of GCP are know as .LOL or .OMG [54], [55].

4) Cerber: As of March 2016, this crypto family has made its newest arrival on the ransomware scene. Since then, Cerber has made a significant impact due to its novelty and nasty behaviors. One of its novel features is its use of a text-to-speech (TTS) module to speak to the victim, which lets the threat read the ransom note loudly [56], [57]. More interestingly, Cerber does not infect users from some particular countries. When a machine is infected, Cerber will check the victim's location. If the computer appears to be from certain countries, Cerber will terminate itself and not encrypt any files. Otherwise, it will perform its malicious activities and encrypt files using AES-256 encryption algorithm [58].

Furthermore, once Cerber is installed, it will name it-self after a random Windows executable. After that, it uses the command C:\Windows\System32\bcdedit.exe"/set{current}safebootnetwork to configure the operating system (Windows) to automatically boot into Safe Mode with Networking on the next reboot. The infected computer will then reboot into Safe Mode with Networking and once the user logs in, Cerber will automatically shut down the computer again and reboot it back into normal mode. Then it will configure itself to run automatically when users log in to Windows, then execute its tasks as one task per minute. Once Cerber executes these tasks, it will show a fake system alert and ask to restart the computer [58].

5) Locky: Locky arrived in late 2015 and early 2016. This nasty and virulent strain of ransomware variants has been propagated widely, mainly through massive spam email campaigns and compromised websites. For example, Locky will attach Word documents, which contain a malicious macro to these emails. Once the macro is run, it will install Locky on the victim's machine.

Even with the increased focus on security in organizations, a few ransomware payloads have made it onto their computers. Hence, many businesses were hit in this onslaught of spam [59]. Since the early Spring of 2016, the emergence of Locky has been counted as one of the most prolific ransomware attacks created to date. Recently, the cybercriminal group behind Locky has begun to use a new downloader, which is known as Rockloader (Downloader.Zirchap), in its spam campaigns. When the victim is infected with Rockloader, it will download and install Locky onto the computer [56].

6) CryptXXX: This family made its first appearance in April 2016. It was circulated widely by the infamous Angler and Neutrino exploit kits. Initial variants of this class used compromised websites to redirect infected computers to the Angler exploit kit and involved Angler first dropping Trojan.Bedep [60] on those computers, which infected them with CryptXXX. This exploit kit was the most popular delivery method used to deliver ransomware variants to potential victims. However, within a few weeks of its release, the Angler exploit kit was dropped and eventually shut down completely. Consequently, it led to a sudden drop off in the activities related to a number of major malware families, including CryptXXX. Since then, this threat is being distributed by the infamous Neutrino and RIG exploit kits [20], [61].

Due to the weak encryption techniques that were used in the initial variants of CryptXXX, security researchers were able to create a decryption tool for compromised computers. However, attackers behind this attack responded quickly and released a newer variant of CryptXXX by employing a better encryption approach. Moreover, the new version of CryptXXX contains a new StillerX credential-stealing module that gives attackers additional capabilities to monetize the attacks. It has also been armed with a sniffing capability that can sniffs out files to encrypt even if they are not locally stored. Thereby, multilayered network and end-user protections remain critical to prevent data exfiltration in case of infection [62].

Furthermore, cybercriminals have continued to refine CryptXXX with more updated features such as scanning for network shares and encrypting them. To find and encrypt shared resources on the network, new variants of this family exhibit scanning activity on the network gateway port 445, which is used for SMB (aka Server Message Block) and primarily associated with Microsoft Windows Domain and Active Directory infrastructure [63]. In addition, CryptXXX has received a major overhaul by its authors and been marked as a top moneymaker for criminals compared to Locky [56].

7) CryptoWall: CryptoWall was first appeared in early 2014 and has the same strategy as many crypto-ransomware types. After the downfall of CryptoLocker, CryptoWall, formerly known as Cryptorbit or CryptoDefense, started to gain no-toriety. It is distributed as fake application updates such as Adobe Reader and Java Runtime Environment. CryptoWall can be facilitated using many typical threat distribution channels, such as pop-up windows if the user visits suspicious websites or opens spam emails. Moreover, CryptoWall 4.0 is now folded into the Nuclear Exploit Kit and can run on both 32-bit and 64-bit systems [64].

In addition, initial variants of CryptoWall used the RSA public key, which is generated on the C&C server, to encrypt crucial files. However, recent variants such as CryptoWall 3.0 use an AES key for file encryption and encrypt the AES key by a public key that is generated on the server. Once a variant infects the computer, it will scan the computer's drives to find files that it can encrypt. CryptoWall variants also scan the locate drive letters on the PC, including network shares, Dropbox mappings, and removable drivers. Furthermore, this nasty type installs malware files either in the %AppData% or %Temp% folders and creates DECRYPT_INSTRUCTION.txt,DECRYPT_INSTRUCTION.html and DECRYPT_INSTRUCTION.url files in directories where the CryptoWall has encrypted data [65], [66].

8) TeslaCrypt: It is a family of crypto-ransomware that was first detected in February 2015. It primarily targets computer games such as the Call of Duty series, World of Warcraft, Minecraft and World of Tanks, and has been widely propagated in mass media as the "curse" of gamers. TeslaCrypt is circulated through exploit kits like Sweet Orange, Angler, and Nuclear, seeking out and encrypting gaming-related files on infected computers using the AES-256-CBC algorithm. Furthermore, authors of this attack keep its strains' encryption schemes updated regularly to steer clear of security researchers. A new update of TeslaCrypt (version 0.4.0) has included new obfuscation and evasion techniques, as well as a new list of file extensions [67], [68].

As of version 0.3.5, TeslaCrypt has the ability to infect not only regular drives that are connected to the computer, but also all the network file resources (shares). This functionality is only available in a few other encryptors like CryptoWall. In addition, after successfully encrypting files, it appends them with different extension names such as .encrypted .ecc, .vvv, .ezz, and .exx [69]. TeslaCrypt is commodity malware, which can be purchased on the underground black market. Cybercriminals pay authors of TeslaCrypt to use its platform and access to various delivery methods such as spam Botnets and exploit kits [70].

9) *Chimera:* Chimera was started as a normal ransomware infection that encrypted local and network files (shares). However, some variant of Chimera (Trojan.Ransomcrypt.V) makes an additional threatening message. After successfully encrypting victims' files, Chimera claims that if the payment is not received, some of the encrypted files (pictures and videos) will be posted on the Internet and other files will remain encrypted. This troublesome type combines its ransomware infection with extortion to make victims pay even when they have backups of their files. Furthermore, it circulates its malicious payloads via Dropbox links in phishing campaigns by sending phishing attacks as job offers, business proposals, and infected email attachments to some employees. Once those links are clicked, Chimera automatically downloads the malware, which will immediately start encrypting targeted files [71], [72].

Despite Chimera's filthy intimidation techniques, it did not perform as expected by its authors. None of the victims' files were ever published on the Internet. However, some other cybercriminal gang managed to steal a significant part of Chimera's source code and build other ransomware classes (Mischa and Petya). Moreover, those thieves decided to ruin the Chimera project by publicly releasing about 3500 of its decryption keys [73].

10) *CryptorBit:* CryptorBit was released in December 2013 and targeted all versions of Windows operating systems. Once it infect a computer, it encrypts any file it scans as opposed to targeting just specific files (as is the case with most types). Therefore, this type can be considered under the indiscriminate encryption class. However, it does not delete users' files. Furthermore, CryptorBit creates a HowDecrypt.txt file and a HowDecrypt.gif in every folder that contains an encrypted file. Those files contain instructions on how to pay the ransom in order to decrypt the files, which can vary from one victim to another. It also installs some software on victims' computers that mines digital crypto-currency (cryptocoin miner). Hence, it allows attackers to utilize the victims' computers' CPU to mine digital coins [74], [75].

In fact, CryptorBit does not encrypt the entire file. It actually corrupts the data header by replacing the first 512 bytes (or 1024 bytes) of the file, which renders the file unusable since programs cannot interpret the corrupted header. In addition, the encrypted bytes will be stored at the end of the original file. Then, it replaces the header with a new 512 bytes. As a result, CryptorBit will effectively corrupt the file because a program that would normally open the infected file would not recognize the header and would not open it. However, this feature makes the data recovery process possible by only repairing the headers and decrypting individual files. This process can be very cumbersome, but it is a valid option in many cases [76]. Table IV shows the percentage of infected users by the top ransomware families as of December 2016 [77].

Table IV: Top 10 Crypto-ransomware families

| Family name | Percentage of infected users |
|---|---|
| CTB-Locker | 25% |
| Locky | 7 % |
| TeslaCrypt | 6.5 % |
| Scatter | 2.85% |
| Cryakl | 2.8 % |
| CryptoWall | 2.3 % |
| Shade | 1.7 % |
| Crysis | 1.1 % |

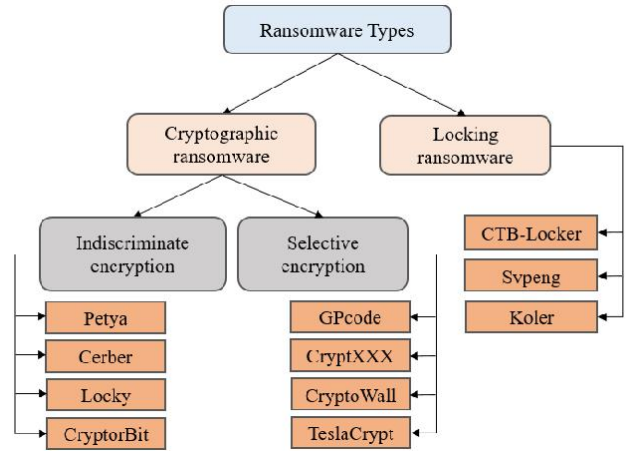

Figure 5: Topmost cryptographic and locking ransomware families

B. Locking Ransomware

Variants of this type deny access to the infected machines without encrypting or deleting files. They can lock out particular resources, such as the mouse or keyboard, in order to ensure limited functionality. Thus, once the ransomware is removed, full access is usually restored. Figure 5 summarizes topmost cryptographic and locking ransomware families.

The following list summarizes the topmost locking ransomware families in the recent years. As attempted in the previous part (cryptographic ransomware), this part aims to identify each form of locking ransomware and how it works:

1) *FLocker:* FLocker is a combination of the words "Frantic" and "Locker," which was appeared in May 2015 and was detected as ANDROIDOS_FLOCKER.A by Trend Micro security experts [78]. These experts have found over 7,000 variants of FLocker, as its author kept updating and creating many variants to improve its routines and avoid detection systems. The common variants of FLocker target mobile devices and smart TVs. For instance, Android.Lockdroid.E is capable of locking Android smart TVs [79]. However, this type does not infect users from particular countries. Therefore, before preforming any malicious behavior, it will check the device's location. If the device is located in certain countries, it will deactivate itself immediately.

After FLocker infects a device, it starts to run its routine after 30 minutes. Then FLocker starts the background service, which requests device admin privileges. If the user denies its request, it will immediately freeze the screen and display a fake system update. Moreover, FLocker runs in the background and communicates with a C&C server, which will deliver a new payload misspelled.apk with an HTML file and a JavaScript interface. The HTML page is able to initiate the APK and take photos of the user by using the JavaScript interface. These photos will be displayed on the ransom page. In addition, device information such as phone number, location, and contacts will be collected by the C&C server and encrypted with a hard-coded AES key and encoded in base64 [80].

2) Koler: This is an Android-based threat that appeared in April 2014. It is promoted as a hidden part of a ransomware campaign and masquerades as fake adult-themed apps. Koler infects Android users from certain countries when they visit suspicious adult-themed websites by asking them to download a fake adult app that would meet their desired contents (animalporn.apk). Unlike PC-based threats, the Koler download is neither silent nor automatic. In other words, the user must confirm the app installation and manually install it. Further, once the app is granted and installed, it will activate Koler and display a law enforcement agency message accusing the user of viewing pornographic contents and demanding a penalty payment [81], [82].

About 48 pornographic websites were infected by Koler. When the user visits one of them, Koler's special controller checks for parameters that must be met such as user's country (location), type of device, and type of browser. If compatible, the promoted letter will be triggered. Moreover, during the installation process, some devices' information will be collected and sent to a C&C server, like the IMEI number. After that, it will lock the device and open a browser page to display a persistent report over the Home screen stating that the device is locked and all its files are encrypted due to some security violations, but this report is fake. It does not encrypt files at all, it does interfere with normal usage of the device to enforce victims to pay their ransoms [83], [84].

Cybercriminals behind Koler refined its scheme such that it has the ability to offer customized attacks and infect either mobile or PC users. When a user visits one of these websites, Koler redirects him to the right infrastructure (central hub) to download the attack. However, the mobile components of this threat have been shut down since July 2014. The C&C server sent uninstall commands to infected devices and deleted the malicious app, whereas the rest of the malicious components for PC users are still active [85].

3) CTB-Locker: The word "CTB-Locker" stands for Curve-Tor-Bitcoin Locker (also known as Critoni.A). The word "Curve" comes from its encryption based, which uses the elliptic curves algorithm. "Tor" is the malicious server that is used to protect the attack anonymity, as most malware variants do. "Bitcoin" refers to the payment method used by most ransomware [86]. CTB-Locker attack was discovered in July 2014 as one of the most dangerous ransomware routines, which was designed to lock victims' computers and deny access to their files. It can be categorized under both crypto and locking classes. Like most types, CTB-Locker comes as a part of a ransomware system in order to bypass the detection techniques. That means it infects a computer with the help of another malicious payload that finds a flaw on the user's computer and utilizes it as an entry point for CTB-Locker attack. In fact, once it infects a PC, it instantly disables any Antivirus software found on that PC [87].

CTB-Locker is mainly delivered through aggressive spam campaigns and email attachments as a zip file such as a UPS exception notification or FedEx delivery failure notification [88]. When the potential victim opens the email, CTB-Locker will ask to download and access the zip file. Once accessed, the attack will be triggered and most files on the system will be encrypted. In fact, this malware is deployed as a binary code that can be executed by just opening the email attach-ments. Furthermore, CTB-Locker will immediately communi-cate with a C&C server and automatically start downloading its components [89].

Like other attacks, authors of CTB-Locker have kept it refined. For instance, in January 2015, they released an updated version of CTB-Locker targeting certain countries including Germany, Italy, the Netherlands, and the USA. Its infection was distributed as a fake fax notification via email attachments. Moreover, to convince victims that the CTB-Locker attack is not invented, this version has introduced a new option called Test Decryption, which allows victims to select and decrypt five different files for free. Another update was released in 2016 to attack websites by encrypting all scripts, documents, databases, and any other important files [90]. Identically, it provides victims with two decryption keys to unlock two random files. In addition, cybercriminals claim that the victims will permanently lose their files if any attempts are made to rid the infected computer of the ransomware [4].

4) Svpeng: Svpeng was first identified in July 2013 by Kaspersky Lab as a Trojan-banker that targets Android devices via MoneyPak [91]. Because it was originally created to steal users' credential information and be aware of the banking apps on the infected devices, it has been modified to perform some ransomware functionality. Hence, it attempts to lock the device and display an accusing message with a fine, reporting that the user has accessed illegal pornographic contents. Within one month of its release, this nasty form was able to infect over 900,000 devices [92].

As other traditional Trojan-bankers for PCs, Svpeng infects mobile devices using some social engineering techniques such as drive-by download, malicious email attachments, fake video players, and Adobe Flash player updates [93]. Furthermore, security researchers identified a distribution campaign that delivers Trojan-banker attacks to Android devices using Google AdSense advertisements. That campaign was launched by authors of Svpeng to infect users when they visit mainstream websites without requiring users to click on the malicious advertisements. In general, cybercriminals keep refining such a threat because mobile attack exploits are often specific to a version of mobile operating system [94].

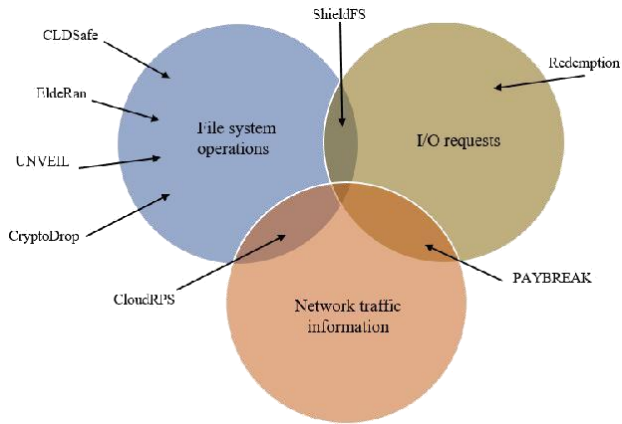

Figure 6: Windows countermeasures based on their traced resources

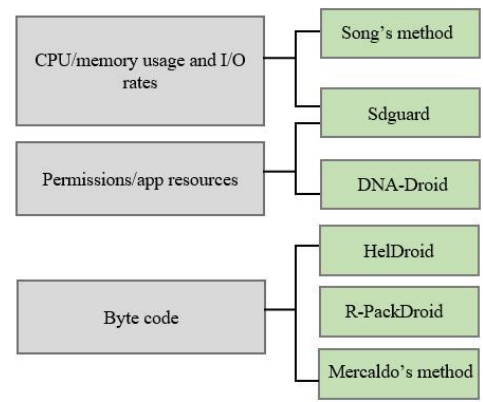

Figure 7: Android countermeasures based on their traced resources

### C. Other Types

There exist some ransomware types that perform neither encryption nor locking activities. Instead, they infect a machine and collect all victims' data. Some other variants are capable of even accessing the webcam and taking pictures. Then, they create a ransom note and threaten the victims that if the ransoms are not paid, those variants will leak the victims' private data. Another notable example of such a type is that some variants (can be listed as scareware) accuse the victims of having engaged in suspicious activity such as viewing pornographic contents or acquiring security violations. Thereby, they threaten victims to pay their ransoms within 24 hours. Otherwise, the scareware alleges that a case against the uses will be sent to trial [95].

## VIII. Security Techniques and Countermeasures

The knowledge about malware's functionality is important for removal. In fact, to be able to completely remove a malware from an infected machine, it is necessary to remove the residues that are left behind by the malicious code and undo modifications that are made to legitimate files. Those residues can be unwanted registry entries, services, or pro-cesses, which require a fully understanding of the malicious code and its operations. However, recent malware tries to evade existing detection techniques by refraining from per-forming any malicious activity once it identifies the analysis environment presence [96]. Therefore, designing a powerful analysis system against malware activities is non-trivial in real-world deployment [7].

There are three types of analysis techniques. These tech-niques are static, dynamic, and hybrid. Static analysis ap-proaches focus on classifying operations based on features that can be collected without running the apps. However, dynamic analysis approaches collect apps' features during run-time. Each technique has its pros and cons. For instance, static analysis can't predict many sophisticated malware. Thereby, both techniques can be integrated to build hybrid detection systems.

After careful selection of all possible methods that are used to detect and prevent ransomware attacks, this section defines these methods and highlights their limitations. It also distills the state-of-the-art in security research and identifies potential research directions for safeguarding billions of users. Note, the scope of this paper has only considered tools that are designed specifically to prevent ransomware attacks. Figure 6 and Figure 7 show the relationships between the security tools covered in this section based on their traced resources in both Windows and Android platforms, respectively.

### A. UNVEIL

This dynamic analysis approach is not at end-user machine. It is designed to detect ransomware based on its interaction with the computer's resources. To successfully amount ransomware attacks and model their behaviors, it creates fake but meaningful resources to allow ransomware variants to tamper with the users' files. In particular, UNVEIL creates an artificial and realistic user environment to execute and monitor ransomware-like process interactions with the file system. Furthermore, it can be used as a complement service on top of other dynamic analysis systems. Also, this approach can be used as a cloud-based malware analysis system and sample sharing. In addition, this system was tested on top of Cuckoo Sandbox, which provides the basic sandboxing services and supports all versions of Windows [7], [97].

In order to address crypto-ransomware variants, authors of UNVEIL created a fake but attractive user environment for the malware to run, which contains real files with valid headers and meaningful names. Then they implemented a kernel-based method to monitor the interactions with the file system. The purposes behind creating this environment are 1) to make the analysis more realistic; and 2) to protect the analysis system from some user environment fingerprinting techniques. After running the malware, this technique collects some process's information such as the time, the process name, the process ID, the request in form of IRP, the argument of a request, and the entropy of data buffer (read or write) [7].

On the other hand, to detect locking-ransomware variants, UNVEIL takes a snapshot of the clear state of the system, then runs the malware. In fact, a ransom note will be dis-played on the screen of the machine once it is infected. Therefore, this detection approach captures another snapshot of the system and measures the dissimilarity score, which is the structural differences between these snapshots. The value of the dissimilarity score is between 0 and 1 computed using Structural Similarity Image Metric (SSIM). The value closes to 1 means that there are huge structural changes on the snapshot's content, and vise versa [7].

The evaluation of UNVEIL was done through two phases. First, it was tested by using about 3500 known samples. Some of those samples are benign apps, but they can perform some ransomware-like behaviors such as DiskCryptor, Winzip, and SDelete. In the second phase, UNVEIL was evaluated by collecting about 148,223 distinct samples, and using 56 UNVEIL-enabled VMs deployed on 8 servers on a Ganeti cluster. In addition, this approach has discovered a new malware family called SilentCrypt [7].

However, the desktop locking variants that use heuristics to look for specific user actions before performing their behaviors (e.g. waiting for multiple reboots or clicks) can evade UNVEIL detection technique. Furthermore, this scheme is not an end-user solution and there was no real end-user interaction involved in its test phase. Another limitation is that attacks can possibly evade the text extraction module by using uncommon ransom words in their notes [7], [97].

### B. CryptoDrop

CryptoDrop is an early-warning detection system that is built to alert users when suspicious applications perform mali-cious activities. This Windows-based approach is a shift from similar initiatives such as the Linux-based Cryptstalker. Cryp-toDrop can work alongside malware detection tools, detecting ransomware-like activities that existing intrusion prevention tools aren't capable of. Moreover, it combines a set of behavior indicators to identify processes that appear to tamper with a large amount of data. Once it recognizes such a ransomware-like attempt behavior, it halts the process and stops it from completing its effort [12], [97].

Authors of CryptoDrop have described three scenarios of crypto-ransomware behaviors based on the variant activities against files. These scenarios are:

1) Overwrite files: malware in this scenario will overwrite files' contents and encrypt them in place.
2) Change files' locations: it changes files' locations and probably rename them before encrypting and dropping them back in their original places.
3) Create new files: this is the most damaging type, which creates new versions of files with encrypted contents and deletes the original ones.

Therefore, CryptoDrop has implemented three primary indicators to inspect malicious executions. One indicator identifies file changes based on modifications to the file byte values. Another indicator measures the similarity between versions of the same file using sdhash function. Last indicator measures the encrypted file's entropy using Shannon Entropy. In addition, CryptoDrop uses two secondary indicators; deletion cases, which is triggered when files are detected after suspicious activity; and file type funneling, which occurs if an app reads a disparate number of files as it writes [12].

Overall, this system monitors changes on the files' con-tents that indicate transformation rather than inspecting ransomware-like attempts and identifying its execution and characteristics. However, CryptoDrop is not intended to stop ransomware attacks from the outset, or attempted to prevent all files from loss. Also, it is unable to investigate the intent of changes. In other words, CryptoDrop cannot determine whether the encryption is done by the user or a ransomware variant. Hence, it may trigger false alerts [12], [97].

### C. PAYBREAK

This is an automated and proactive defense system designed to combat against crypto-ransomware threats and keep the user files safe, specifically against WannaCry attacks. PayBreak intercepts and stores all the cryptographic materials that can be used during the attack in a key vault. That means, it controls the file encryption process by monitoring the session keys that are used on the victim's machine. Furthermore, this Windows-based approach observes and can hold the use of these keys so that it can decrypt files that would otherwise only be restored by paying the ransom [34], [97].

In fact, modern ransomware uses hybrid cryptosystem, which is a combination of symmetric and asymmetric keys to encrypt files. A symmetric key is used to encrypt each targeted file. Then a public key will be generated to encrypt these symmetric keys while the decryption private key is held by the ransomware attacker. When an attack infects a computer, PayBreak records and stores the symmetric keys before they get encrypted by the public key. With that, the victim can retrieve the ransom keys and decrypt infected files [34].

In addition, the effectiveness of PayBreak was evaluated by running 107 ransomware samples of 20 distinct families in a controlled environment. Those samples are collected by using a system designed by authors of PayBreak called Real-time Automation to Discover, Delete, and Alert of Ransomware (RADDAR). However, this approach was only able to recover files encrypted by 12 family variants. Moreover, it failed to determine samples that use environment fingerprinting tech-niques, as well as variants that use heuristics to look for specific user actions before performing their behaviors. Also, it performs its protection task with a performance overhead due to its dynamic hooking techniques and the key escrow mechanism [34].

### D. Redemption

Redemption is a real-time data-loss-fee protection framework designed to make the OS more resilient to ransomware threats. This end-user approach monitors apps' I/O request patterns on a per-process basis to indicate the ransomware-like behaviors. When an I/O request pattern is identified for a possible ransomware activity, the offending process will be terminated. Therefore, the data can be restored. However, Redemption requires some modifications of the OS to maintain a transparent buffer for I/O storage [97].

Furthermore, the performance of this approach is based on two main components. The first component contains a constructed behavior abstraction of a large class of characterizations of current ransomware attacks. To determine whether the process exhibits a malicious behavior or not, Redemption applies the results of a long-term dynamic analysis to the binary objects. If the process matches the abstract model, it will be labeled as malicious. The second component contains a transparent buffer utilization mechanism used to redirect requests while tracking the write contents in order to restore all infected files [97].

The evaluation of this system was run on 677 ransomware samples from 29 families. Redemption was capable of enhancing protection against the collected attacks, achieving a very high detection rate in such a successful application-transparent manner. However, it fails to distinguish between benign apps that exhibit some ransomware-like behaviors. Also, if it happens that the malice score of a process is lower than the detection threshold, it would not be discriminated. In addition, this framework is unable to detect social engineering techniques, which can be used by cybercriminals to shut it down unless they are detected by other security solutions running on the same computer [97].

### E. CLDSafe

This file backup system is a cloud based designed against ransomware to back up user's files automatically and keep shadow copies safe in order to provide secure restoration when a machine is infected. CLDSafe saves copies of user's files in a cloud storage system after it measures the similarity score between the new file on user's machine and the old version stored in the cloud storage using a separate place called secret area. The reason of using this cloud storage system is that commercial cloud storages provide limited spaces, where users cannot keep all backups in one storage. Also, some ransomware forms are capable of infecting backup shadows if they are synchronized with the infected machines such as CryptoLocker attacks. Thus, the secret area in CLDSafe is needed to overcome autonomous synchronization [98].

Furthermore, to investigate ransomware infections, the similarity score is computed using a context triggered piecewise hash technique called ssdeep [99]. Once an infection is discovered, CLDSafe allows authenticated users to restore backup files using challenge-response mechanism in the secret area. This approach backs up files only when there are changes made to them. Moreover, when the similarity score is low, which indicates changes on the new files, CLDSafe copies the old files to the secret area in order to restore them in a secure manner. In addition, CLDSafe is capable of detecting and blocking storage-consuming attacks such as denial of service attack. The Storage consumption occurs when the resources are fully occupied. However, CLDSafe cloud storage system only backs up a file when there is a notable changes, which does not cause much memory or CPU consumption [98].

CLDSafe id evaluated using 210 file modification cases as follows. The fist 200 cases were ordinary cases from four types of data file. These types are .pptx, .xlsx, .docx, and .hwp. Each type has five samples and each sample is measured ten times. The remaining ten cases were ransomware files infected by CryptoLocker. The similarity threshold of this approach is set to 80 to lower the storage overhead to 51% and keep the overall performance efficient if compared to commercial cloud storages [98]. However, CLDSafe would failed against some ransomware variants that only encrypt a part of the data file. A notable example of those variants is CryptorBit, which corrupts the data header by replacing the first 512 bytes or 1024 bytes. When the encrypted part is less or equal to 20% of the total data size, the similarity score will be above the threshold. Therefore, CLDSafe considers it as an ordinary case.

### F. CloudRPS

This ransomware prevention approach uses a cloud analysis technique to collect and analyze device information aiming to detect abnormal behaviors and minimize the possibility of early infections. In particular, CloudRPS collects data from an integrated cloud system and analyzes these data to investigate ransomware intrusions [100]. This approach consists of six components as follows:

1) Server Monitor: This component consists of condition manager to check the server's status, activity status, and whitelisting; and reporter to send the results to the next component.
2) File Monitor: It is used to monitor files. It consists of threat prevention, which helps to update security patches of the device's programs; folder management to automatically hide shared folders and allow authorized access only; initial classifier to monitor file's operation; network control to control network communications; and reporter to share analysis results.
3) Network Monitor: It monitors and analyzes the traffic and reports of user's device. The collected elements in this component include network information, network signature, traffic frequency, and whitelisting.
4) Ransomware Inspector: It is used to monitor the user device changes and manage the device state and information. This component contains some prevention factors such as OS update, network, data backup, server, and file check. Moreover, these factors are synchronized in real time through the Cloud system.

5) Analyzer: This component performs a hybrid analysis. It also contains a reporter that reports results to the classifier.
6) Classifier: Its task is to categorize the transmitted data that is received from the network monitor, the files monitor, the server monitor, and the analyzer based on the threat level and store the result in a database.

CloudRPS detects many servers', networks', and files' ran-somware attacks by monitoring filesystem operations [100]. However, as most cloud-based detection systems, this ap-proach is limited in terms of storage spaces. Also, synchroniz-ing files can cause time consumption. Therefore, the infection might not be prevented.

### G. ShieldFS

ShieldFS is a Windows-based add-on driver designed to make the filesystem immune to ransomware attacks. This approach contains a protection layer that can be dynamically toggled for each running process and performed as a copy-on-write mechanism. Furthermore, it contains a set of adaptive models that copy the system activity over time. These models are updated by a monitoring component that investigates low-level and I/O filesystem requests. Once a process violates these profiled models, its operation will be determined as a malicious behavior [101].

ShieldFS was designed after analyzing a large amount of low-level and I/O filesystem requests collected from a set of benign apps. Its detection technique was built based on the following parameters: the combined analysis of entropy in write operations; frequency of read, write, and folder-listing operations; dispersion of per-file writes; fraction of files renamed; and the file-type usage statistics. Moreover, it makes a decision within each tick, which is the clock of ShieldFS. Ticks are triggered based on the percentage of files accessed. All these parameters are considered to create an automated detection models that can indicate ransomware processes at run-time, as well as the use of cryptographic primitives by scanning the memory of any malicious process and searching for traces of the typical block cipher key schedules [101].

The detection technique of ShieldFS was applied in a real-time, self-healing virtual filesystem that shadows the write operations. That means that when a malicious process alters a file, the filesystem will present the original, mirrored copy to the user space apps as a shadowing mechanism that can be activated and deactivated dynamically depending on the detec-tion logic. These copies will be deleted when the process has been cleared as benign. In addition, ShieldFS was evaluated by running 688 ransomware samples collected from 11 distinct families in real-world working conditions. It was able to achieve a very high accuracy score by detecting the malicious activity at runtime and transparently recovering all the original files. However, authors of ShieldFS have mentioned a few limitations [101]. The most critical issues are highlighted as follows:

1) Susceptibility to Targeted Evasion: If the thresholds of the classifiers and the value of the parameter T (which determines how often ShieldFS should create shadowed copies of the files) are known by the malware, it could attempt to perform mimicry attacks by encrypting a few files and remain below the thresholds for T hours. In this case ShieldFS would not indicate as a malicious process.
2) Multiprocess Malware: Ransomware can fork its malicious code by injecting many benign apps, each of which performs a small part. Such an attack can evade the detection technique of this system.
3) Tampering with the Kernel: ShieldFS runs in a privileged kernel mode. However, administrator privileged processes can shut down ShieldFS services when the machine is booted.

### H. EldeRan

This is a machine learning system designed to dynamically analyze ransomware and classify them based on their early attempts. It monitors behaviors that can be performed by apps in their installation phase by inspecting for ransomware-like behaviors and characteristic signs before they infect victims. Furthermore, EldeRan identifies relevant dynamic features that can be used to detect ransomware. Then, it employs a machine learning classifier to classify each installed app such that it can provide detection without relying on classical heuristics or signature-based techniques [8].

Additionally, EldeRan identifies the most significant ran-somware dynamic features and creates signatures for new variants. More precisely, it performs its dynamic analysis in a sandboxed environment that contains two datasets: a ransomware dataset of 582 samples from 11 distinct families and a benign dataset of 942 goodware samples. Within each sample, EldeRan retrieves and analyzes the following features:

Windows API calls and native functions tracing
    Registry key operations (open, read, write, and delete)
    File system operations (open, read, write, and delete)
    Directory operations (the enumeration and creation oper-ations performed on directories)
    Dropped files (set of files dropped by an application during installation)
Strings (embedded in the binary)

Except the Strings, these features are collected while dy-namically analyzing apps. The machine learning technique in EldeRan consists of two stages: 1) feature selection, and 2) classification. In the first stage, it applies a feature selection algorithm (Mutual Information criterion) to select the most relevant dynamic features, whereas in the second stage, a Regularized Logistic Regression classifier is applied to the matrices that contain these features to distinguish between ransomware and goodware applications [8].

The experimental results of this approach have proven its efficiency and effectiveness. Moreover, the accuracy of the Regularized Logistic Regression has been evaluated by comparing it with other machine learning classifiers such as Support Vector Machine (SVM) and Naive Bayes. It outperforms Naive Bayes and is competitive with respect to the SVM. Also, it shows how dynamic analysis can be used to utilize the set of characteristic features that are common across ransomware families at run-time. However, EldeRan fails to extract features of some samples where they are silent for sometime or variants that use heuristics to look for specific user actions before they perform their malicious activities [5], [8].

### I. HelDroid

This automated system combines static taint analysis with lightweight symbolic execution to find possible paths that indicate ransomware-like behaviors. It considers "building blocks" that are typically needed to build a ransomware app. HelDroid has the ability to identify apps that attempt to lock or encrypt the device without the user's awareness, as well as ransom notes that are displayed on the screen. Additionally, this approach uses a learning-based, "natural language processing (OpenNLP)" technique that recognizes menacing phrases to detect threatening aspects of ransomware [43].

HelDroid analyzes each Android app APK file to determine whether it is a ransomware or benign app. In its detection process, it employs three independent detectors that are executed simultaneously. Each detector investigates a specific indicator of typical ransomware-like behavior. The first detector is called "Threatening Text Detector," which uses text classification to detect coercion attempts and threatening messages. The result of this classifier depends on the other detectors. If it is positive, but the others are not, the sample will be labeled as scareware. The second and third detectors are the "Encryption Detector" and the "Locking Detector." If the app performs either action, one of these detectors will be triggered. Therefore, the application will be labeled as a ransomware sample [43].

Furthermore, the deterministic decision criteria of HelDroid is based on static analysis. Although most of its analysis process is static, the Threatening Text Detector executes the code in case there is no ransom note found in the files, which supports off-band text (such as messages loaded from a remote server). In addition, it considers the presence of the ransom note as mandatory for a ransomware attack to reach their intention. Thus, if the Threatening Text Detector is not triggered, the app will be considered a benign sample [43].

Overall, HelDroid is implemented and tested on active Android ransomware samples. Its detection features are para-metric and can be adaptable to future families. Hence, it works without requiring certain samples of any ransomware family beforehand except a small portion of sentences obtained from ransomware sample threatening notes. It successfully exhibited nearly zero false positives and detected all the ransomware samples. Although it fails to detect ransomware samples that use evasion mechanisms or embedding cryptographic primi-tives, it is capable of recognizing unknown samples. However, ransomware notes that are written in foreign languages are not supported by the Threatening Text Detector. As a result, re-training the NLP classifier with foreign language ransom texts is required. Moreover, HelDroid should be integrated into the device OS in a trusted domain. This crucial integration allows the system to block malicious code actions. To solve such a problem, both encryption and locking indicators must have high priority [43].

### J. Sdguard

This proposed solution implements fine-grain permission control based on Linux Discretionary Access Control (DAC) mechanism to detect crypto-ransomware attempts. Sdguard consists of two main components. The first component is the activity stack monitor, which monitors Android activity stack. The second component is the I/O log analyzer and the access control list, which allows users to grant certain permissions to the installed applications. When an app creates a file, Sdguard uses FUSE file system (Filesystem in Userspace) to modify the owner and group of the file according to the UID and GID of the app. Moreover, the FUSE daemon contains a permission checker module to verify permissions and file's UID before determine an access [102].

Another service that FUSE daemon provides is an I/O recorder, which records all I/O operations to external storage, then write log to a file. This log file is parsed to investigate malicious behavior existence by the I/O log analyzer. On the other hand, the activity stack monitor observes an activity that is located on top of the stack. Once it finds an activity on top of the stack for long time and this activity does not belong to the Android lock-screen service, it will be determined as a ransomware activity attempting to lock the user screen. Therefore, this activity is killed to eliminate the threat [102].

### K. R-PackDroid

R-PackDroid [103] is a machine learning approach designed to detect Android ransomware based on extracted API package information. This static detection system is used to label inspected applications as one of three classes; either ransomware, malware, or trusted app. To identify the app's classes, this system uses three phases as follows:

1) Preprocessing: In this stage, the inspected app's Dalvik bytecode is statically analyzed to determine the packages of all APIs.
2) Feature Extraction: Features are selected based on the occurrences of each API package, which identify the vector numbers that are used in the next stage.
3) Classification: Extracted data from the aforementioned phase is forwarded to a trained mathematical function that statically classifies and label the inspected app. In particular, random forest classifications are employed.

R-PackDroid was evaluated and tested by using data that are different than the data used to train the system. These datasets were collected from different resources such as HelDroid [43], Drebin [104], and VirusTotal [105]. Moreover, an open source crawler was used to download trusted apps from Google Play market. In addition, R-PackDroid exhibits a high detection rate with a very good performances. However, as it entirely performs as a static analysis system, R-PackDroid is subject to some limitations such as suspicious payloads that are dynamically loaded at run time or fully encrypted classes. Also, this system was not tested against obfuscated applications. Therefore, it uses VirusTotal service to confirm its classification results and reduce its false positive rate [103].

L. DNA-Droid

DNA-Droid is a hybrid analysis system used to detect ransomware applications in Android platform. This framework utilizes features and deep neural network to discriminate between suspicious and goodware samples. DNA-Droid aims to detect ransomware actions in early stage by extracting app's static features and label it. If the app is labeled as suspicious, DNA-Droid applies dynamic analysis to monitor run-time behaviors. The inspected app will be terminated once its operations match with collection of malicious behaviors (DNAs), which are produced during the training phase. In addition, DNA-Droid uses Binary and Multiple Sequence Alignment techniques to profile ransomware families [106].

In the static stage, DNA-Droid includes three modules to evaluate miscellaneous aspects of Android APK files as follows:
1) Text Classification Module (TCM): To detect ran-som notes, DNA-Droid uses Natural Language Toolkit (NLTK) to perform linguistic analysis. It extracts strings from the apk file and constructs ransom words in dif-ferent classes such as encryption, locking, threatening, pornography and money. For each class, TCM calculates the similarity score to indicate the existence of each class in the inspected apk.
2) Image Classification Module (ICM): Similar to TCM, ICM extracts images form the sample apk file to match it with different logos classified into different categories such as banks, police, and low enforcements. To measure the similarity score, ICM uses the Structural Similarity Index Measure algorithm (SSIM) and reports the number of collected images. Moreover, to detect porn pictures, ICM uses a skin color model to classify them to nude or non-nude pictures.
3) API calls and permissions Module (APM): It extracts permissions and APIs used by the inspected apk file.

Furthermore, DNA-Droid uses Deep Auto Encoder to im-prove detection rate and classification performance. In the dynamic analysis stage, DNA-Droid executes the app using an emulator and collects some information such as system sequence and API calls. If the app's API sequences match with any of the DNA, the app will be terminated [106].

M. Song et al. Method

Many ransomware attacks in mobile platforms are not new to the existence. Cybercriminals keep refining threats because mobile attack exploits are often specific to a version of operating system [94]. Therefore, Song et al. proposed a method that can prevent modified ransomware attacks in Android system. This solution uses statical methods based on CPU and memory usage as well as I/O rates to monitor execution operations and files events and specify abnormal operations. Once a suspicious operation is identified, it will be terminated and reported to the user and stored to a database for future investigation. This method is implemented in the source code of Android. Thus, it can detect modified ransomware patterns [46]. Furthermore, this method is designed with three modules as follows:
1) Configuration Module: It is the basic step of this detection technique. Its job is to specify the files location that are targeted by the attack and register them in a watch list table. The location of these files is called priority protection area (PPA). Also, this module registers user's feedback on the inspected process into the database. If the use reports it as ransomware, it will be stored and automatically deleted. Otherwise, the process will keep running even if it is detected again.
2) Monitoring Module: It is responsible for monitoring the PPA and the process. This module consists of two parts. The fist part monitors the status of the file event such as reading, writing, and delete. On the other hand, the second part monitors the process information such as memory and I/O count. In addition, this module handles malicious behaviors stored in the database by stopping and deleting them once they are detected.
3) Processing Module: It stops the monitors process if it is reported as suspicious and warns the user about the risk behind it. Moreover, processing module removes the malicious apps and deletes any process belongs to it.

However, this method reports the detection result to the user in order to either allow the process or deny it, which harms the user's files if granted accidentally. Also, users must be knowledgeable about ransomware attacks and how they perform in order to provide the right decision to this method. Otherwise, their files would be infected. These limitations can be fixed by completely killing the process and then report it to the user with a recommendation to uninstall the corresponding app. Another solution can be done through applying a dynamic analysis on new/updated applications installed on the device to detect ransomware apps in their early stage.

N. Mercaldo et al. Method

Another method was developed by Mercaldo et al. [41] based on formal methods to identify the ransomware instruc-tions inside the malware code. This framework is structured by following three processes:
1) Formal Model Construction: It extracts and parse the formal methods from the app's Java bytecode. This process uses a custom parser based on the Apache Commons Bytecode Engineering Library (BCEL) to parse the bytecode of classes and JAR files. To pro-duce the formal methods from the parsed bytecode, this approach exploited the Calculus of Communicating Systems (CCS).

Table V: Covered tools' summary

| Platform | Name of Tool | Supplementary Technique | Targeted Families | Notable Limitations |
|---|---|---|---|---|
| | UNVEIL | Cuckoo Sandbox, Structural Similarity Image Metric (SSIM), Windows Minifilter Driver framework | CryptoWall, CTB-Locker, CryptoLocker, Tox, Reveton, CrypVault, VirLock, CoinVault, Filecoder, TeslaCrypt, Tobfy, and Urausy | Heuristics variants, Not an end-user solution |
| | CryptoDrop | File utility program, sdhash, Shannon Entropy | Virlock, CTB-Locker, MBL Advisory, CryptoLocker, Trojan-FUE2, CryptoWall, CryptoDefense, Pgpcoder, CryptoFortress, CryptoTorLocker2015, Filecoder, Xorist, and PoshCoder | Do not prevent all files from loss, Cannot determine the encryption source |
| | ShieldFS | Virtual filesystem with shadowing mechanism | CryptoWall, TeslaCrypt, CryptoDefense, Crowti, and Critroni | Susceptibility to evasion, fails against DoS, multiprocess malware, and tampering with the kernel attacks |
| | PAYBREAK | Real-time Automation to Discover, Delete, and Alert of Ransomware (RADDAR), fuzzy function, Microsoft Research's Detours library, Cuckoo Sandbox | Almalocker, Cerber, Chimera, CryptoFortress, CryptoLocker, CryptoWall, CrypWall, Tox, GPcode, Locky, SamSam, and Thor Locky | Can't recover all infected files, fails against fingerprinting techniques and heuristics variants |
| | CLDSafe | Fuzzy hashing (ssdeep context triggered piecewise hashing), cloud storage system, challenge-response mechanism | This tool was tested against CryptoLocker variants only | Needs a separate place to restore files in order to overcome autonomous synchronization, fails against variants that encrypt 20% or less of the file |
| | CloudRPS | integrated with a cloud system, unknown classifier | not specified samples | limited storage space, large files cause time consumption. Therefore infections are not prevented |
| | Redemption | Microsoft Reparse Points, Behavioral Detection and Notification Module, Malice Score Calculation Function (MSC), Recursive Feature Elimination (RFE) | CryptoDefense, CryptoLocker, CryptoFortress, CTB-Locker, Tox, CryptoWall, WannaCry, Jigsaw, SilentCrypt, Filecoder, TorrentLocker, TeslaCrypt, CryptXXX, GPcode, PoshCoder, MBL Advisory, Virlock, ZeroLocker, Locky, CoinVault, CrypVault, Crowti, CryptMIC, DirtyDecrypt, HDDCryptor, Xorist, Petya, and Critroni | Fails to distinguish benign apps that exhibit some ransomware-like behaviors, unable to detect social engineering techniques |
| | EldeRan | Mutual Information criterion and Regularized Logistic Regression classifier | CryptoWall, Critroni, CryptoLocker, Kollah, Matsnu, Pgpcoder, Reveton, Kovter, Locker, TeslaCrypt, and Trojan-Ransom | Can't extract features from silent samples and heuristics variants |
| | Heldroid | Cybozu open-source library, Google Translator, Stop-word Project, Optical Character Recognition (OCR) software, natural language processing (NLP) supervised classifier | Koler, Simplocker, Svpeng, New Simplocker, ScarePackage, other unknown types | Threatening Text Detector supports only few languages, must have high priority in the device OS |
| | Song et al. Method | Modifications applied to the Android source code | Authors used some sample to test their approach and did not specified targeted ransomware families | Depends on user's feedback, users must be aware of the attack to provide the right decision |
| | DNA-Droid | Deep Auto Encoder, Binary and Multiple Sequence Alignment, Structural Similarity Image Metric (SSIM), Natural Language Toolkit (NLTK) | Crosate, Koler, Locker, FakeInst, Spy, Simplocker, Jagonca, Torec | Produce a high false positive rate. Therefore; it uses supplementary techniques to improve detection and classification performance |
| | Sdguard | Linux Discretionary Access Control (DAC), Filesystem in Userspace (FUSE) file system | Proposed solution has not specified targeted ransomware families | Users have to be knowledgeable in order to set up specific access rules, it dealers some permissions to root the device |
| | R-PackDroid | Scikit-Learn ML library, open source crawler, VirusTotal | Svpeng, Simplocker, New Simplocker, Koler, ScarePackage, and some unknown samples | Use VirusTotal to reduce false positive rate, unable to detect suspicious behaviors that are loaded at run time or fully encrypted classes, was not tested against obfuscated apps |
| | Mercaldo et al. Method | Apache Commons Bytecode Engineering Library (BCEL), mu-calculus, Concurrency Workbench of New Century (CWB-NC), Calculus of Communicating Systems (CCS) | Locker, Koler, fbilocker, scarepackage, FakeInstaller, Plankton, DroidKungFu, GinMaster, BaseBridge, Adrd, Kmin, Geinimi, DroidDream, Opfake | Evaluated against general malware solutions, evaluation did not include better solutions such as McAfee and Kaspersky |
| No More Ransom | | provides many security solutions for Teslacrypt, Chimera, Marsjoke, Coinvault, Shade, Rakhni, WildFire, and Rannoh ransomware attacks | | |

2) **Temporal Logic Properties Construction:** This process defines the ransomware distinctive characteristics and behaviors. In order to do that, a few ransomware samples were manually inspected to specify their properties and write them as a set, which is used to find the ransomware payload in the bytecode. Moreover, mu-calculus [107] is used to express the behavioral properties.

3) **Ransomware Family Detection:** In this process, a formal verification environment called Concurrency Workbench of New Century (CWB-NC) [108] is applied to rec-ognize the ransomware type based on its set of logic properties.

Furthermore, using bytecode in malware detection system has many benefits such as independence of the source pro-gramming language, identify malware without decompilation, easiness of parsing low-level code, and independence from ob-fuscation [41]. In addition, authors of this method encouraged using it against ransomware threats based on their experiment results. However, this method was evaluated against some general malware solutions that are not specifically designed against ransomware attacks. Authors also stated that those solutions are the top ten signature-based antimalware. In Fact, authors did not include better solutions that are ranked as one of the top ten signature-based antimalware such as McAfee, Kaspersky, TotalAV, and Norton.

### O. No More Ransom

This project was established by a group of IT security vendors and law enforcement organizations in July 2016 to disrupt ransomware cybercriminal businesses. It is led by Europol, the European Union's law enforcement agency, including Intel Security and Kaspersky Lab [109], and has since added 13 new law enforcement agencies all around Europe. This effort provides many security solutions like prevention advice, investigation assistance, wealth of information on ransomware, and decryption tools. For instance, they provide decryption tools for Teslacrypt, Chimera, Marsjoke, Coinvault, Shade, Rakhni, WildFire, and Rannoh ransomware attacks. According to a security report generated by McAfee Labs in December 2016, No More Ransom has allowed victims to avoid paying more than US$1.48 million to cybercriminals. Furthermore, its website has received more than 24.5 million visitors since it was launched (about 400,000 visitors every day) [20], [110].

Due to the availability of ransomware code and creative derivatives, which are provided as Ransomware-as-a-service in dark markets, most security companies find it easier to avoid the threat than to fight against it. That means when a machine is infected by a ransomware attack, this project aims to educate users about how ransomware performs and what suitable countermeasures can be used to prevent such an attack. The more security vendors involved in this initiative, the better the results provided can be. Hence, it is open to other public and private parties to join. In addition, with such an initiative and with more parties as well as development and release of ransomware prevention techniques, the volume and effectiveness of ransomware attacks will be reduced.

However, this project provides solutions only for some forms of ransomware. Therefore, cybercriminals can keep deploying threats and attacking users by using other variants and target-ing other countries [20], [110].

## IX. USER POLICIES AND RECOMMENDATIONS

Ransomware threats have recently occupied a big part of the threat landscape. Business organizations and individuals who have been infected with ransomware attacks may decide to merely accept the situation and pay ransoms without any further investigation. Hence, it is vital to educate users about ransomware threats and encourage them to adopt best solutions. Malware such as scareware can take on the appearance of ransomware attacks to distract victims while the real malicious activity is active to perform some data theft attack. This section aims to educate both organizations and individuals to safeguard their environments and themselves against ransomware attacks as follows:

### A. Backups

Attackers are capable of pushing their malicious activities to millions of users. They often lock/encrypt resources using strong and unbreakable techniques. Therefore, one of the key pillars of combating attacks of ransomware is backing up valuable data. However, as illustrated in this study, some ransomware attacks are even capable of encrypting or deleting backups. Also, restored access to encrypted data by paying the ransom is not always guaranteed. Hence, backing up data in external storage is a must, and it should not be a replacement for a robust security strategy. It is highly recommended that users validate recovery points and update backups. Also, they are encouraged to have backup storage in a location that is only accessible by a secure mode system.

### B. Network Scan and Traffic Monitoring

Network administrators must run a habitual test on their environments to fix existing bugs that cybercriminals may use. They should keep security software updated and rely on strong indicators of trust. Also, they must perform a full network scan to investigate malware infections. If any compromised computer is identified, it should be isolated from the network until it is fully cleaned and restored.

Furthermore, monitoring network traffic is a network management process that is applied to study the network communications and ensure the normality of the network performance and security. The process of monitoring network traffic in-corporates network sniffing and packet capturing techniques to review each incoming and outgoing packet. Therefore, to mitigate ransomware attacks, scanning the network and monitoring its traffic play a key role in identifying suspicious operations such as encryption and locking activities.

### C. Phishing Emails

Ransomware attacks are spread using multiple infection vectors. However, cybercriminals know the majority of people who are aware of the capability of ransomware attacks.

Therefore, they continue to refine more advanced tactics. They usually use spear-phishing emails in the first stage of attacking victims. End-user training for organizations and individuals would reduce the risk of malicious emails from being opened and spread in the first place. Users are highly advised to immediately delete any suspicious emails they receive.

In particular, emails from unrecognized senders, which may contain links and/or attachments that prompt users to enable macros, can be used for legitimate purposes. For instance, attackers can use malicious macros to run malware through Microsoft Office documents and perform automated tasks. Thus, Microsoft has mitigated this infection by disabling macros from loading in its documents by default. However, attackers can use other ways to convince users to run it such as using social engineering techniques.

### D. General Recommendations

A successful attack on a targeted organization can potentially compromise thousands of computers, ending up with a massive operational disruption and serious damages to their revenue. Therefore, they must be aware of the threats posed by ransomware attacks. Moreover, they must have multilayered security solutions to minimize the chance of malware infec-tions and become more competent. Most importantly, security administrators must be aware of the problem and the way it spreads. The following list contains a number of policies and procedures that would minimize the success of ransomware attacks if followed:

> Keep system and software patches updated and verify if they are applied successfully. Many security enhance-ments are released frequently to mitigate discovered vulnerabilities.
> For systems and devices that cannot be patched, risks can be mitigated by leveraging app whitelisting, which prevents the execution of unapproved programs.
> These systems and devices must be isolated from the network using firewalls and intrusion detection system (IDS). Also, it is highly recommended to disable unnecessary services and ports to reduce the possibility of exploiting entry points.
> If possible, do not store sensitive data on local disks. It should be stored on secure network drives. That will reduce the downtime of infected systems, which can be recovered by simply re-imaging them. Also, be aware of your critical data's location and the methods that might be used to infiltrate it.
> Unwanted programs and traffic must be blocked. If Tor is not needed, block the app and its traffic on the network. As most ransomware attacks use Tor to perform their activities, blocking Tor will often stop the encryption process.
> Use virtual infrastructure for critical systems that are isolated from the rest of the network.
> If users need to connect to untrusted resources, they should use secure VPN middleware.

## X. CONCLUSION

This article has done a systematic review of the terms related to ransomware attacks and summarized behavioral descriptions of the most common families in the Windows and Android platforms. Ransomware variants are the most prevalent attacks of today. Practically, they aim at user's victimization. With the effective extortion schemes that support financial malware development, cybercriminals are getting more sophisticated in the way they craft their malware. This paper has briefly sum-marized the background of ransomware attack and distilled the state-of-the-art of recent research. Its scope has covered the major existing security efforts, which can provide users with rich information and security solutions to achieve their objectives. Security countermeasures are carefully selected from miscellaneous aspects to meet most research needs in this particular type of attack. Finally, this research can benefit the security community as well as researchers to further safeguard organizations and individuals.

### ACKNOWLEDGMENT

This work was supported in part by NSF under grants CNS-1460897, DGE-1623713, DGE-1723707 and the Michigan Space Grant. Any opinions, findings, and conclusions or recommendations expressed in this material are those of the authors and do not necessarily reflect the views of the NSF.